\documentclass[pre,aps,preprint,nofootinbib,amsmath,floatfix]{revtex4}

\usepackage{graphicx}

\usepackage{dcolumn}
\usepackage{bm}
\usepackage{amssymb}

\newcommand{\tw}{\textwidth}

\begin{document}

\title{Oscillating behavior of a compartmental model with retarded noisy
  dynamic infection rate}

\author{Michael Bestehorn}
\affiliation{Brandenburgische Technische Universit\"at Cottbus-Senftenberg,
  Institut f\"ur Physik, Erich-Weinert-Stra{\ss}e 1, 03046 Cottbus, Germany\\
  bestehorn@b-tu.de}

\author{Thomas M. Michelitsch}
\affiliation{Sorbonne Universit\'e, Institut Jean le Rond d'Alembert, 
CNRS UMR 7190, 4 place Jussieu, 75252 Paris cedex 05, France\\  
thomas.michelitsch@sorbonne-universite.fr}

\begin{abstract}
Our study is based on an epidemiological compartmental model, the SIRS model.
In the SIRS model, each individual is in one of the states susceptible (S), infected(I) or
recovered (R), depending on its state of health. In compartment R, an individual is
assumed to stay immune within a finite time interval only and then transfers back to the
S compartment. We extend the model and allow for a feedback control of the infection rate by
mitigation measures which are related to the number of infections.
A finite response time of the feedback mechanism is supposed that changes the
low-dimensional SIRS model into an infinite-dimensional set of integro-differential 
(delay-differential) equations. It turns out that the retarded feedback
renders the originally stable endemic equilibrium of SIRS (stable focus) into an
unstable focus if the delay exceeds a certain critical value. Nonlinear solutions
show persistent regular oscillations of the number of infected and susceptible individuals.
In the last part we include noise effects from the environment and allow for a
fluctuating infection rate. This results in multiplicative noise terms and our model
turns into a set of stochastic nonlinear integro-differential equations. Numerical
solutions reveal an irregular behavior of repeated disease outbreaks in the form of infection
waves with a variety of frequencies and amplitudes.\\[2ex]
\end{abstract}


\maketitle

\section{Introduction}

Mathematical modeling of epidemic dynamics goes back to the seminal work of Kermack and
McKendrick [1927] where the nowadays called 'SIR model' was introduced, an
acronym from (${\bf S}=$ susceptible, ${\bf I}=$ infected, ${\bf R}=$
recovered). The basic SIR model and its various extensions (for a review see
\cite{anderson,mart}) are also called
'compartmental models' since they divide the individuals into several compartments depending
on their state of health. It turned out that the features of infectious diseases such as
measles, mumps, and rubella could to a certain extend be captured by such simple models.
A huge field has emerged to describe epidemic spreading in the framework of random walks in
complex networks
\cite{Satorras-Vespigniany2001,BesRiascosMichel2020,RiascosMateos2021,besmi} and
(generalized) fractional dynamics
\cite{MetzlerKlafter2000,SandevChechkinMetzler2021,mipo}, just to name but a few.
A model based on a small world network is discussed in
\cite{small,small1} and proved to be superiour to SIR (or SEIR) models
comparing its results with data provided by the SARS outbreak in mthe year 2003.

Time periodic outbreaks have been noticed for a long time in the dynamics of several
diseases and were already stated in 1929 by Soper in a model for the time evolution of
measles cases \cite{soper}.

The SIR model and most of its extensions are not able to describe sustained oscillations
but rather account for a single outbreak caused by the instability of the disease free
state. In the long time limit, the endemic equilibrium is reached where the fractions of the
population in the different compartments attain constant values. In the language
of dynamical systems this behavior is known as a heteroclinic orbit, connecting
in phase space an unstable fixed point (healthy state) with a stable one (endemic state).
In the following we shall consider an extended SIRS model where the time of immunity
is finite and in the endemic equilibrium a nonzero number of infected individuals
remains present so that the disease never become extinct completely.
In the original SIR or SIRS models, the interplay between infected and susceptible
individuals is inspired by the dynamics of the even older predator-prey systems
\cite{lotka} and is expressed in the form of a simple bilinear term
$\beta_0\,I(t)\,S(t)$, where $I(t)$ and $S(t)$ are the number of infected and
susceptible individuals and $\beta_0$ is the constant probability of infection at each
contact (infection rate). The predator (infected) 'catches' the prey (susceptible) by
infection.

Other work \cite{liu} considered a nonlinear infection rate according to
\begin{equation}\label{nonin}
 W = \beta(j)\,I(t)\,S(t) = \frac{\beta_0\,I^m(t)\,S(t)}{1+\alpha I^n(t)} \ ,
\end{equation}
and obtain limit cycle solutions for certain parameters $m=n\ge2,\;\alpha>0$.
Tang et al. [2008] studied the case $m=n=2$ and found a weak focus and the
existence of two limit cycles.
For $m\ge n$, $W$ is a monotonically increasing function of $I$ that saturates for $m=n$. 
For $m=n=1$ the dynamics is qualitatively the same than for the standard SIR model
and sustained oscillations cannot be observed.
The denominator $1+\alpha I^n$ accounts for  mitigation measures against the epidemics
that naturally increase with increasing $I$. We only note that for the case $n>m$
the interaction $W$ has a maximum at a certain infection number.
For such nonmonotonic behavior it was shown in \cite{xiao} that the dynamics in the
long time limit approaches a stable fixed point as for the original SIR or SIRS models.

In the present paper we want to stay as close as possible at the standard SIRS model
and will not consider additional limit cycle solutions obtained for $m=n\ge2$ or
non-monotonic functions $W$ of $I$. We therefore
study a SIRS model with the most simple nonlinear interaction of
the form (\ref{nonin}) with $m=n=1$. Taking the recent Covid epidemic as an example,
such a feedback control is realized by certain containment measures like social
distancing, hygiene rules, or wearing masks. Normally these measures take effect after a
certain retardation that can be on the same time scale or even much longer than the time of
recovery. The time delay can be either distributed or singular. In our model we
generalize (\ref{nonin}) by replacing $I(t)$ in the feedback by a memory integral
as follows
\begin{equation}\label{nonin1}
W = \frac{\beta_0\,I(t)\,S(t)}{1+\alpha I_d(t)} \quad\mbox{with}\quad
I_d(t) = \int_{-\infty}^{t}K(t-\tau)I(\tau){\rm d}\tau \ .
\end{equation}
For the distributed case, $K(t)$ is a given causal normalized probability density
function (PDF) introducing memory into the model.
The singular time delay is a special case with $K(t-\tau)=\delta(t-\tau-\tau_0)$ with
Dirac's $\delta$-function.

Delay or memory terms were introduced in epidemiological models by many other
researchers, for an overview see \cite{rihan}. In a previous work we considered a SIRS
model with a retarded transition from the R to the S compartment, reflecting the rather
long time of decay of immunity \cite{besmi}. From the mathematical point of view,
the presence of a delay term
in an ordinary differential equation makes a low-dimensional system infinitely
dimensional and may allow for the occurrence of periodic, quasi-periodic or even chaotic
behavior, rendering the dynamics much more complex \cite{hutch,mack,bes1}. 
Memory effects introduced by finite incubation periods, delayed infectiousness and the
distribution of the recovery period are considered in an upcoming paper
\cite{Basnarkov-et-al2021}.

The main focus of the present paper is to analyze an epidemiological model that allows for
persistent periodic outbreaks of the disease, in contrast to the standard SIR or SIRS
models that show an asymptotically constant endemic equilibrium.
To simulate environment fluctuations, noise terms are added
that may have a significant influence of the amplitude and period of the oscillations.

The paper is organized as follows: After introducing the modified SIRS model with retarded
feedback control in part II, we perform in part III a local linear stability analysis
close to endemic equilibrium. Threshold parameters are determined for which the endemic
equilibrium becomes oscillatory (Hopf) unstable. The existence of a Hopf unstable
endemic equilibrium is crucial for the occurrence of sustainable periodic outbreaks.
For the kernel in (\ref{nonin1})
we consider a $\delta$-function and especially an Erlang PDF which contains two free
parameters and turned out to be flexible enough to capture real-life situations
\cite{besmi}. In Part IV, fully nonlinear solutions for these cases
are given above threshold and the occurrence of persistent oscillations
around the endemic equilibrium is observed.
Finally, noise terms are introduced into the infection rate, accounting for a fluctuating
environment \cite{cai}. It is demonstrated that these terms are responsible for
much more irregular oscillations, showing the typical behavior known from real-life data.

\section{Model}

\subsection{SIRS model with feedback control}

\noindent
Let $S$, $I$, $R$ be the number of susceptible, infected, and recovered individuals,
respectively and $N=S+I+R$ their total number. Assuming a bilinear interaction
between susceptible and infected individuals, the SIRS model has the form 
\begin{subequations} \label{sirorg}
\begin{align}
  \frac{dS}{dt} & = -\frac{\beta_0}{N}\,I\,S + \nu\, R \\
  \frac{dI}{dt} & = \frac{\beta_0}{N}\,I\,S -\gamma\,I \\
  \frac{dR}{dt} & = \gamma\,I  - \nu\, R \ ,
\end{align}
\end{subequations}
where $\beta_0$ is the average number of contacts per individual per time, multiplied
by the probability of disease infection between a susceptible and an infectious
individual, and $1/\gamma$ is the average
time of being infectious or the time of recovery. The parameter
$\nu$ is the immunity loss rate and accounts for a finite life time of
immunity $1/\nu$. For $\nu=0$ the standard SIR model \cite{kermack}
is recovered.

Since no birth or death processes are considered in (\ref{sirorg}), the 
total number of individuals $N$ is constant in time. 
Instead of the numbers $S,I,R$ we introduce the fractions
$s(t), j(t),r(t) \in [0,1]$
\begin{equation}\label{sir_def}
s(t)=\frac{S(t)}{N},\quad j(t)=\frac{I(t)}{N},\quad r(t)=\frac{R(t)}{N}
\end{equation}
and obtain from (\ref{sirorg})
\begin{subequations} \label{sir}
\begin{align}
  \frac{ds}{dt} & = -\beta_0\,j\,s + \nu\, r \label{sira} \\
  \frac{dj}{dt} & = \beta_0\,j\,s -\gamma\,j \label{sirb} \\
  \frac{dr}{dt} & = \gamma\,j  - \nu\, r \label{sirc} \ ,
\end{align}
\end{subequations}
Scaling the time with $\gamma$ and considering $r+j+s=1$,
the system (\ref{sir}) is reduced to the nondimensional form
\begin{subequations} \label{sir1}
\begin{align}
  \frac{ds}{dt} & = -R_0(j)\,j\,s + \mu\,(1-j-s) \label{sir1a} \\
  \frac{dj}{dt} & = R_0(j)\,j\,s - j \label{sir1b} \ ,
\end{align}
\end{subequations}
with $\mu=\nu/\gamma$ and
\[ R_0(j) = \frac{\beta(j)}{\gamma} \]
as the basic reproduction number. From here we allow for an infection number dependent
infection rate $\beta(j)$ to model feedback control by mitigation measures, see
eq. (\ref{nonin}). For arbitrary $R_0(j)$, the system (\ref{sir1}) has a fixed point
\begin{equation}\label{fx1}
j_h = 0,\quad s_h = 1 \ , 
\end{equation}
corresponding to the disease free state and becoming unstable for $R_0(0)>1$.
For constant $R_0>1$, the other fixed point
\begin{equation}\label{fx2}
j_e = \frac{\mu(R_0-1)}{R_0(\mu+1)},  \quad s_e = \frac{1}{R_0} 
\end{equation}
denotes the endemic equilibrium and is unconditionally stable. Note that for
$\mu=0$, (\ref{fx2}) turns into $j_e=0,\ s_e=1-r_e$, where $r_e$ depends on the initial
conditions and on the dynamics. 

If containment measures take effect the infection rate $\beta$ will decrease.
Since the strength of the measures normally increases with the number of
infected individuals, it is nearby to assume a certain dependence
$\beta=\beta_0/f(j)$ or
\begin{equation}\label{rdyn}
R_0(j) = \frac{r_0}{f(j)}, \qquad r_0=\beta_0/\gamma
\end{equation}
and $f(j)\ge1$ as a monotonically increasing function of $j$ with $f(0)=1$.
The endemic equilibrium is now found from the solution of
\begin{equation}\label{fix2}
j_e\,(1+\mu) + \frac{\mu f(j_e)}{r_0} - \mu = 0
\end{equation}
and depends on $f$. Taking the most simple form (\ref{nonin1})
\begin{equation}\label{fsim}
f(j) = 1+\alpha j,\qquad \alpha\ge0 \ ,
\end{equation}
eq. (\ref{fix2}) is linear in $j_e$ and
\begin{equation}\label{fx2a}
  j_e = \frac{\mu(r_0-1)}{r_0(\mu+1)+\alpha\mu},
  \quad s_e = \frac{1}{r_0}\,\left(1+\alpha j_e\right) \ .
\end{equation}
The infection number of the endemic equilibrium is monotonically decreasing with
increasing $\alpha$ due to the
containment measures. It exists again only for $r_0>1$ where it is proved to be always
stable. for $r_0>1+\mu/4+O(\mu^2)$ the endemic equilibrium is a stable focus.
For $r_0\gg1$, $j_e$ reaches the saturation value $\mu/(1+\mu)$ independent
on $f$.

\subsection{Retarded infection rate control}

\noindent
The containment measures are not instantaneously coupled to the number of infected but
follow them rather with a certain time delay. To include this issue, we introduce the causal
probability density function (PDF) $K(\tau)$ from which the finite time of delay between
cause and effect is drawn. Instead of (\ref{rdyn}) we may formulate
\begin{equation}\label{rdyn1}
  R_0(j) = \frac{r_0}{f(j_d(t))}
\end{equation}
with the retarded infection
\begin{equation}\label{rdyn1a}
j_d(t) = \int_{-\infty}^{t}K(t-\tau)j(\tau){\rm d}\tau \ .
\end{equation}
The delay-time  PDF is normalized,
\[ \int_{0}^{\infty}K(t){\rm d}t = 1. \]
The complete model reads
\begin{subequations} \label{sird}
\begin{align}
  \frac{ds}{dt} & = -\frac{r_0\,j\,s}{f(j_d)} + \mu\,(1-j-s) \label{sirda} \\[2ex]
  \frac{dj}{dt} & =\frac{r_0\,j\,s}{f(j_d)} - j \label{sirdb} \ .
\end{align}
\end{subequations}
Its solutions are defined by the control parameters $r_0,\ \mu$ and depend
also on the form of $f(j)$ and $K(t)$. Due to the memory term, the initial
conditions have to be extended to
\[ s(0),\quad j(t),\ -\infty<t\le 0 \]
if the memory is infinite. In practice however, the memory has a finite length $t_0$ where
$K(t>t_0)\rightarrow0$. Then it is sufficient to integrate in (\ref{rdyn1a})
from $t-t_0$ to $t$ and fix the initial conditions for $j(t)$ on the stripe $-t_0<t\le 0$.

\section{Stability of the endemic equilibrium}

\subsection{Characteristic equation}

For the rest of the paper we assume $f$ given as (\ref{fsim}).
To investigate the stability of the endemic equilibrium, we insert
\[ s = s_e + u\,\mbox{e}^{\lambda t},\quad j = j_e + v\,\mbox{e}^{\lambda t} \]
into (\ref{sird}) and linearize with respect to $(u,v)$. The solvability condition
reads
\begin{equation}\label{solv}
  P(\lambda) =
  a\lambda^2 + \lambda\,\left(r_0 +\alpha\tilde K(\lambda) +a\mu\right)
  +r_0\,(1+\mu) + \mu\alpha\,\tilde K(\lambda) = 0 \ ,
\end{equation}
where $\tilde K(\lambda)=\int_0^\infty\exp(-\lambda t)K(t)dt$ stands for the Laplace
transform of $K(t)$ and $a=1+1/j_e$
with $j_e$ from (\ref{fx2a}). Since $\alpha,a,\mu,r_0>0$ and $\tilde K(0)=1$,
there exists no real valued $\lambda=0$ as solution of (\ref{solv}). As a consequence,
the endemic equilibrium (\ref{fx2a}) can only become unstable due to an oscillatory (Hopf)
instability.

\subsection{$\delta$-kernel}

\begin{figure}[!ht]

\centerline{\includegraphics[width=0.5\tw]{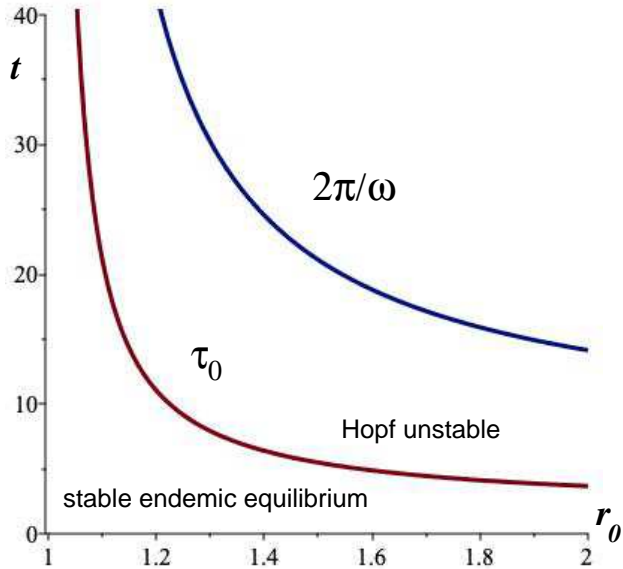}}  

\caption{\label{fig3} $\tau_0$ and $2\pi/\omega$ according to
(\ref{solvom},\ref{solvtau})
  as a function of $r_0$ for $\alpha=50$ and $\mu=0.1$. Time in units of the recovery
  time $1/\gamma$. Above the red line the endemic equilibrium is oscillatory unstable.}

\end{figure}

Now we need to define the memory kernel. The most simple form is
\[ K(t)=\delta(t-\tau_0) \]
where $\tau_0$ is the singular delay time between cause (high incidence) and effect
(measures become active)
and $\tilde K(\lambda) = \exp(-\lambda\tau_0)$. Inserting $\lambda=i\omega$, (\ref{solv})
is separated into real and imaginary parts and a quadratic equation for the
Hopf frequency $\omega^2$ can be derived:
\begin{equation}\label{solvom}
a^2\omega^4 +\omega^2\,\left(r_0^2+a^2\mu^2-\alpha^2-2r_0 a\right) +
r_0^2\,(1+\mu)^2 -\alpha^2\mu^2 = 0
\end{equation}
from where $\omega$ is determined from the larger root. Finally, $\tau_0$ follows from
\begin{equation}\label{solvtau}
\tau_0 = \frac{1}{\omega}\,
\arccos\left(\frac{-r_0(\mu(1+\mu)+\omega^2)}{\alpha(\mu^2+\omega^2)}\right) \ .
\end{equation}
Fig.\ref{fig3} shows $\tau_0$ and the time period $2\pi/\omega$ for
which the fixed point $j_e,s_e$ becomes oscillatory unstable as a function of $r_0$ for
fixed $\alpha=50$ and $\mu=0.1$.

\subsection{Erlang kernel}

\noindent
Another candidate for the kernel which is able to capture a variety of behaviors
is the so called Erlang distribution (also called gamma-distribution)
which has the form  
\begin{equation}
\label{Erlang}
K_{\eta,\xi}(t) =  \frac{\xi^{\eta}t^{\eta-1}}{\Gamma(\eta)} e^{-\xi t}, \quad \eta >0,
\quad \xi >0,\quad t\geq 0 \ ,
\end{equation}
where the index $\eta$ may take any positive (including non-integer) values and
$\Gamma(\eta)$ denotes the Euler Gamma-function which recovers the standard factorial
$\Gamma(\eta+1)= \eta!$ when $\eta\in\mathbb{N}_0$. 

The Erlang distribution (\ref{Erlang}) contains two parameters
$\eta,\;\xi>0$ which may take any positive values.
The Erlang PDF has the Laplace transform 
\begin{equation}
\label{erlangfourier}
{\hat K}_{\eta,\xi}(\lambda) = \int_{-\infty}^{\infty} e^{-\lambda t} \Theta(t) K_{\eta
}(t) {\rm d}t = 
\frac{\xi^{\eta}}{(\xi + \lambda)^{\eta}} \ ,
\end{equation}
where $\Theta(t)$ indicates the Heaviside unit step function which comes into play by
causality. The Erlang distribution is able to capture a variety of pertinent situations.
For $\eta=1$ the exponential PDF is recovered. Further the two extreme cases
of a globally sharp time of immunity $\tau_0=\eta/\xi$ as well as a broadly scattered
distribution can be described. A sharp expected immunity life time $\tau_0$ is
obtained by the limit 
\begin{equation}
\label{feature_flexible}
\lim_{\xi \to \infty} K_{\xi\tau_0,\xi}(t) =\delta(t-\tau_0).
\end{equation}
where $\tau_0$ is constant.
This feature can easily be seen by performing this limit in Fourier space, replacing
in (\ref{erlangfourier}) the Laplace variable with $\lambda=i\omega$, thus
\[\lim_{\xi \to \infty} (1+i\omega/\xi)^{-\xi \tau_0} = e^{-i\omega\tau_0} =
  \int_{-\infty}^{\infty}e^{-i\omega t}\delta(t-\tau_0){\rm d}t \]
which is the Fourier transform of Dirac's $\delta$-distribution
$\delta(t-\tau_0)$. A broadly scattered distribution is obtained for
${\hat K}_{\eta,\xi}(\lambda) \to 0+$ for $\lambda>0$ whereas
${\hat K}_{\infty,\xi}(0) =1$ (normalization) is maintained. In this situation the
parameters are chosen such that the Erlang variance is diverging
\[ \langle (\Delta t)^2 \rangle = \frac{\eta}{\xi^2} \to \infty \ . \]
\begin{figure}[!ht]

\centerline{\includegraphics[width=0.9\tw]{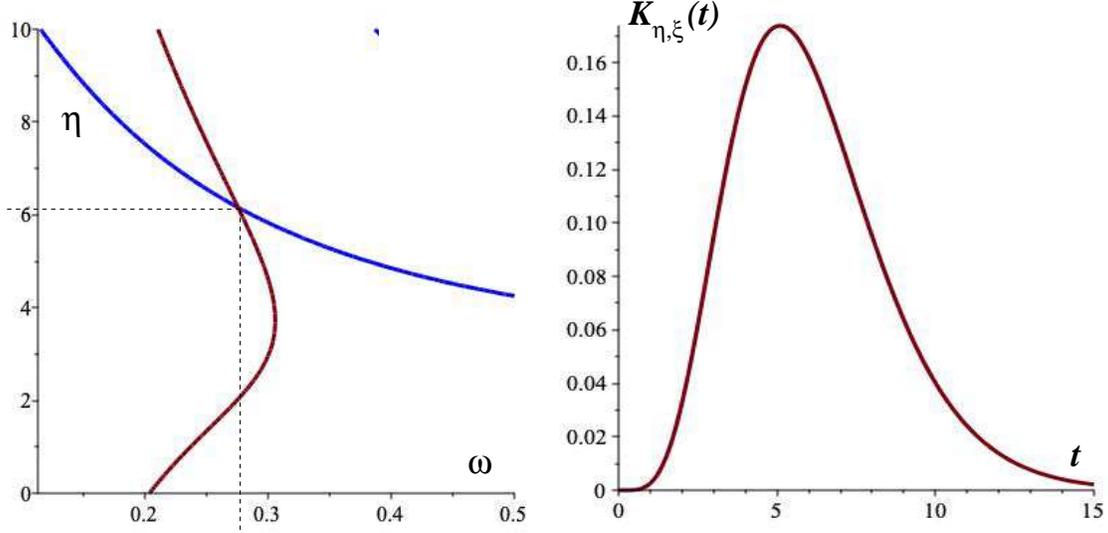}} 

\caption{\label{fig3a} Left: the zeros of imaginary part (blue) and real part of
  $P(i\omega)$ intersect at $\omega\approx0.28$ and $\eta\approx6.1$ for
  $\xi=1,\ r_0=1.6$ and other parameters as in fig.\ref{fig3}. Right: Erlang
  distribution for $\eta=6.1,\ \xi=1$.}

\end{figure}
For $0<\eta \leq 1$ the Erlang distribution is completely monotonic, for $\eta >1$
it possesses a maximum at $t_m=(\eta-1)/\xi$.
The Erlang PDF has a finite mean (expected response time of measures)
$\langle t \rangle =\int_0^{\infty} t K_{\eta,\xi}(t) = \eta/\xi$, i.e.
large $\eta$ and small $\xi$ increase the response time.

Inserting (\ref{erlangfourier}) into (\ref{solv}), an analytic solution for $\omega$
is no longer accessible. Instead we propose a graphical solution by plotting the
zero contours of real and imaginary parts of $P(i\omega)$ in the $(\omega,\eta)$ plane
and looking for their intersections (fig.\ref{fig3a}). Thus, for fixed $r_0$ and $\xi$
a minimal value of $\eta$ for the instability of the endemic state as well as the
Hopf frequency can be determined.

At $r_0=1.6$ and $\xi=1$ we see from fig.\ref{fig3a} a minimal value of $\eta\approx6.1$. In
this case the Erlang distribution has its maximum at $t_m\approx5.1$.

\section{Numerical solutions}

\subsection{Deterministic model\label{sec4a}}

\subsubsection{$\delta$-kernel}

\begin{figure}[!ht]

  \centerline{\includegraphics[width=\tw]{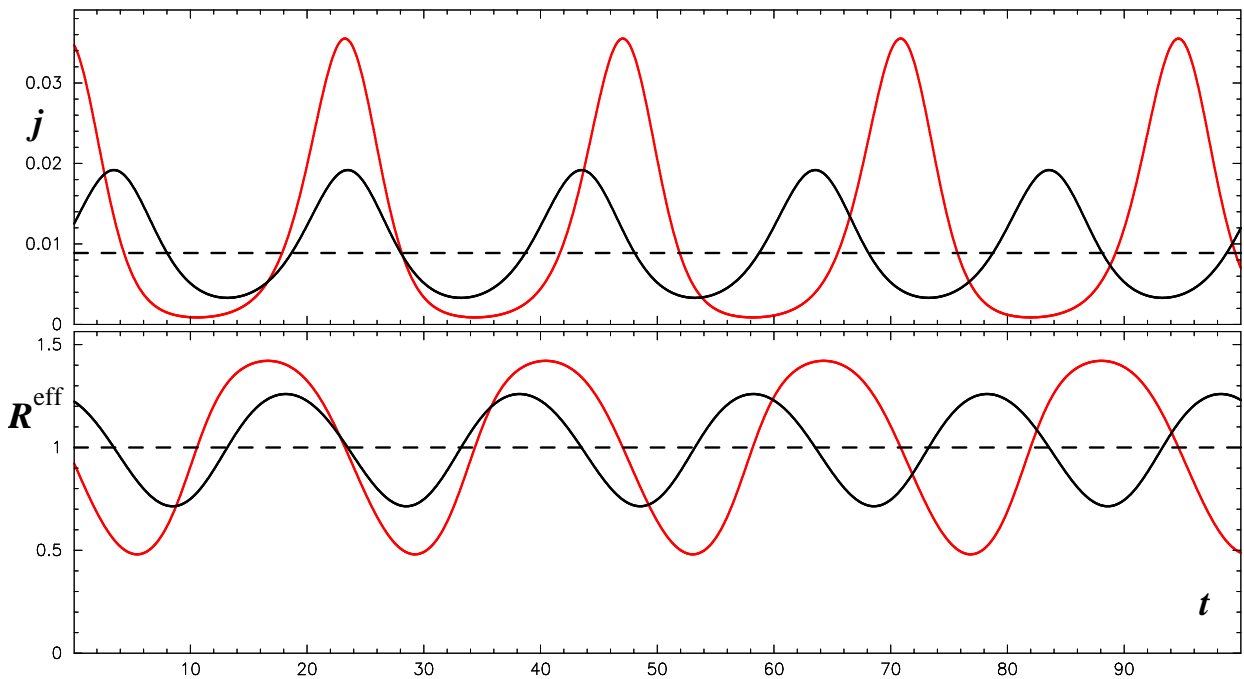}}  

\caption{\label{fig4} Top: $j(t)$ over time for (\ref{sird}) with delta-kernel
  $K(t)=\delta(t-\tau_0)$, dashed line is the endemic equilibrium $j_e$. Bottom:
  effective reproduction number. Parameters as in fig.\ref{fig3}, $r_0=1.6$,
  $\tau_0=5.2$ (black) and $\tau_0=6.2$ (red). Time in units of the recovery
  time $1/\gamma$.}
  
\end{figure}

The system (\ref{sird}) is solved numerically by a 4th order Runge-Kutta method with
fixed time step $\delta t=0.001$ \cite{besbuch}. For the $\delta$-kernel, the last
$n=\tau_0/\delta t$ values of $j$ are stored to compute the delay term. Fig.\ref{fig4}
shows the number of infectious and the actual effective reproduction number
\begin{equation}\label{reff}
R^{\mbox{\small eff}}(t) = \frac{r_0\,s(t)}{1+\alpha j_d(t)}
\end{equation}
over time $t$. For $r_0=1.6$ the endemic equilibrium becomes unstable for
$\tau_0> 4.9$ with the Hopf frequency $\omega=0.33$. If $\tau_0$ is increased, the
oscillations become more and more anharmonic, their frequency decreases and their
amplitude increases significantly, together with the mean values of $j$. 
We find
  \[ <j> = j_e = 0.009 \ (\tau_0<4.9),\quad
   <j> = 0.01 \ (\tau_0=5.2),\quad
   <j> = 0.012 \ (\tau_0=6.2) \ . \]

\subsubsection{Erlang kernel}

\begin{figure}[!ht]

  \centerline{\includegraphics[width=\tw]{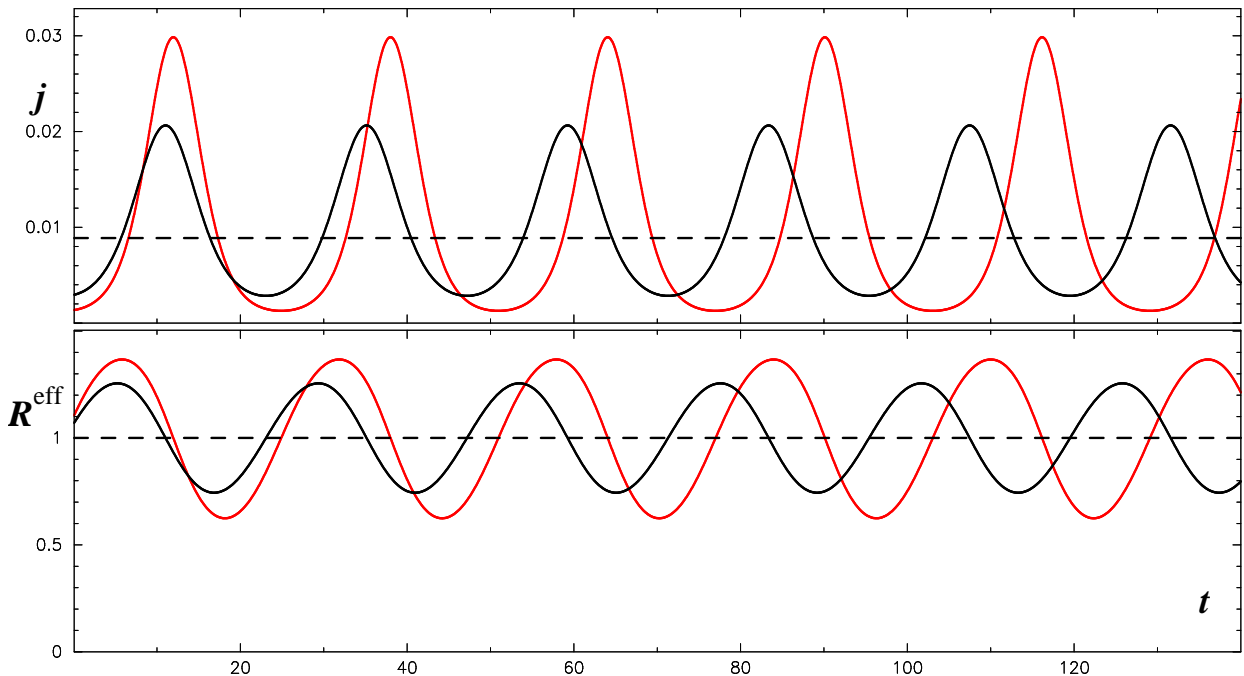}} 

  \caption{\label{fig5} Same as fig.\ref{fig4} but now for the Erlang distribution
    with $\xi=1,\ \eta=6.5$ (black), and $\eta=7.0$ (red).}

\end{figure}

Taking the Erlang distribution, the memory integral (\ref{rdyn1a}) must be approximated
with a finite lower limit
\begin{equation}\label{rdyn1a1}
j_d(t) = \int_{t-t_0}^{t}K(t-\tau)j(\tau){\rm d}\tau \ .
\end{equation}
and numerically evaluated by a sum over the last $n=t_0/\delta t$ time steps.
We chose $t_0=5\,t_m$ where $t_m$ denotes the maximum of $K(t)$, resulting in an error
in the order of $K(t_0)/K(t_m)\approx2\cdot 10^{-6}$.
For the largest $\eta=7$ we have $n=30\,000$.
Fig.\ref{fig5} shows the number of infectious and the actual effective reproduction number
for this case, again for the parameters of (\ref{fig4}) for $\eta=6.5$ and $\eta=7.0$.
From the linear theory onset of oscillations is expected at $\eta\approx6.2$, compare
fig.\ref{fig3a}. The results are at least qualitatively similar to those of the
$\delta$-kernel.
This is due to the fact that the Erlang distribution for
$\eta\approx6$ has a pronounced and rather sharp maximum, cmp. fig.\ref{fig3a}, left
frame. On the other hand, a monotonically decreasing kernel would not lead to
an oscillatory instability.

\subsection{Stochastic model}

There exist plenty of possibilities to extend the model considering noisy perturbations
coming from the environment. A nearby assumption is that of a fluctuating infection
rate which was studied for a SIRS model without memory and therefore without an
oscillatory instability in \cite{cai}. To this end we replace $r_0$ in (\ref{sird}) by
\[ r_f(t) = r_0\,(1+\sigma\,\xi(t)) \]
where $\xi(t)$ is a Gaussian distributed random variable (white noise) with
\[ <\xi(t)>=0,\qquad <\xi(t)\xi(t')> = \delta(t-t') \]
and $\sigma$ denotes the noise intensity. Thus, the stochastic model reads now
\begin{subequations} \label{sirdst}
\begin{align}
  ds & = \left[-\frac{r_0\,j\,s}{f(j_d)} + \mu\,(1-j-s)\right]\,dt
  -\frac{\sigma\, r_0\,j\,s}{f(j_d)}\,dW \label{sirdsta} \\[2ex]
  dj & =\left[\frac{r_0\,j\,s}{f(j_d)} - j\right]\,dt
  +\frac{\sigma\, r_0\,j\,s}{f(j_d)}\,dW \label{sirdstb} \ .
\end{align}
\end{subequations}
where $dW$ is the one-dimensional Wiener process \cite{gardiner} with
\[ dW =\xi(t)\,dt \ . \]

A numerical realization of (\ref{sirdst}) applying a stochastic Euler forward method
(Euler-Maruyama scheme) \cite{kloeden} with time step $\delta t$ reads
\begin{subequations} \label{sirdstn}
\begin{align}
  s_{k+1} & = s_k + \left[-\frac{r_0\,j_k\,s_k}{f(j_{dk})} + \mu\,(1-j_k-s_k)\right]\,\delta t
  -\frac{\sigma\, r_0\,j_k\,s_k}{f(j_{dk})}\,z_k\sqrt{\delta t} \label{sirdstna} \\[2ex]
  j_{k+1} & =j_k + \left[\frac{r_0\,j_k\,s_k}{f(j_{dk})} - j_k\right]\,\delta t
  +\frac{\sigma\, r_0\,j_k\,s_k}{f(j_{dk})}\,z_k\sqrt{\delta t} \label{sirdstnb} \ .
\end{align}
\end{subequations}
where $j_k =j(k\delta t),\ j_{dk} =j_d(k\delta t),\ s_k=s(k\delta t)$ and $z_k$ is a
Gaussian or Bernoulli distributed uncorrelated random variable with mean zero and
variance one,
\begin{equation}\label{zk}
<z_k>=0,\qquad <z_kz_\ell> = \delta_{k\ell}
\end{equation}
and  $\delta_{k\ell}$ denotes the Kronecker symbol. For $\delta t\rightarrow0$, the scheme
(\ref{sirdstn}) converges to the It\^o stochastic ODE system (\ref{sirdst}).

\subsubsection{$\delta$-kernel}

\begin{figure}[!ht]

  \centerline{\includegraphics[width=\tw]{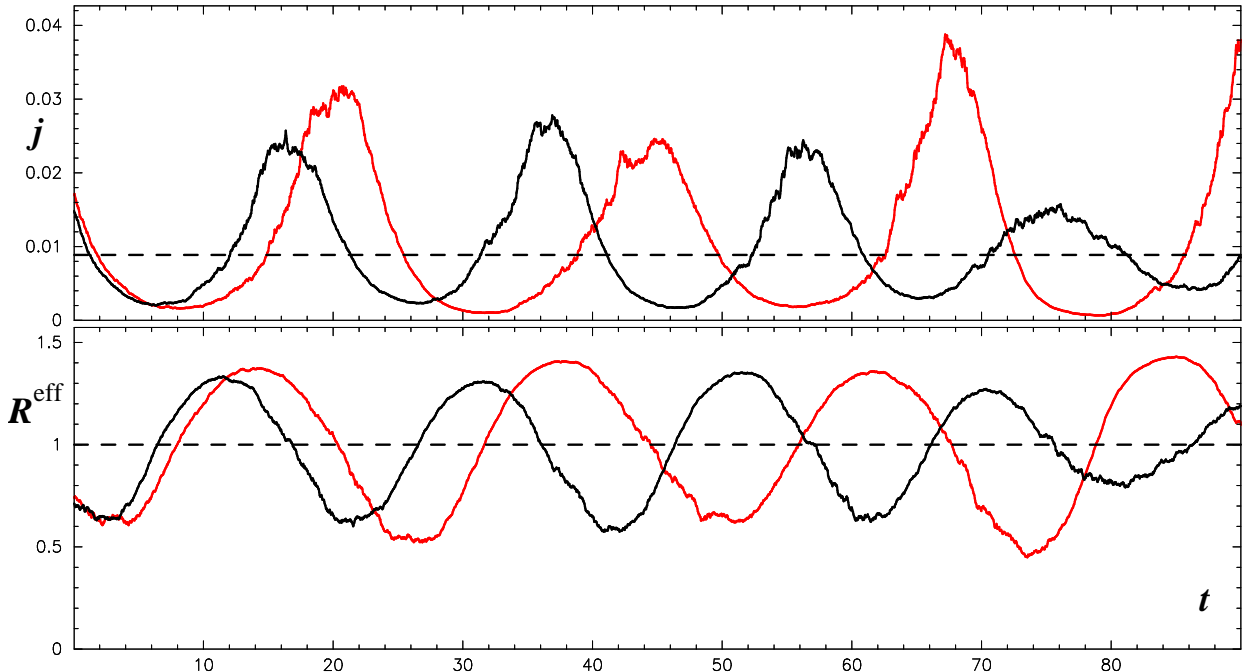}}  

  \caption{\label{fig6} Infection number and $R^{\mbox{\small eff}}$ for the stochastic
    $\delta$-kernel model with $\sigma=0.075$, other parameters as in fig.\ref{fig4}.}

\end{figure}

We repeat the simulations of sect.\ref{sec4a}, first with the $\delta$-kernel, including
now fluctuations. System (\ref{sirdstn}) is integrated numerically. For accuracy reasons
we treated the deterministic part again by a 4th order Runge-Kutta scheme with
$\delta t=10^{-3}$. The random variable $z_k$ is computed by an equally distributed
series $z_k=\pm1$ with probability 1/2, fulfilling (\ref{zk}). The result for $\sigma=0.075$
is shown in fig.\ref{fig6}. A main influence of the noise terms can be seen on the
amplitudes of the oscillations. Contrary to the series of fig.\ref{fig4} there is now
no distinct difference between the amplitudes of $\tau_0=5.2$ and $\tau_0=6.2$. The main
frequency decreases with increasing delay time for both cases.

\subsubsection{Erlang-kernel}

\begin{figure}[!ht]

  \centerline{\includegraphics[width=\tw]{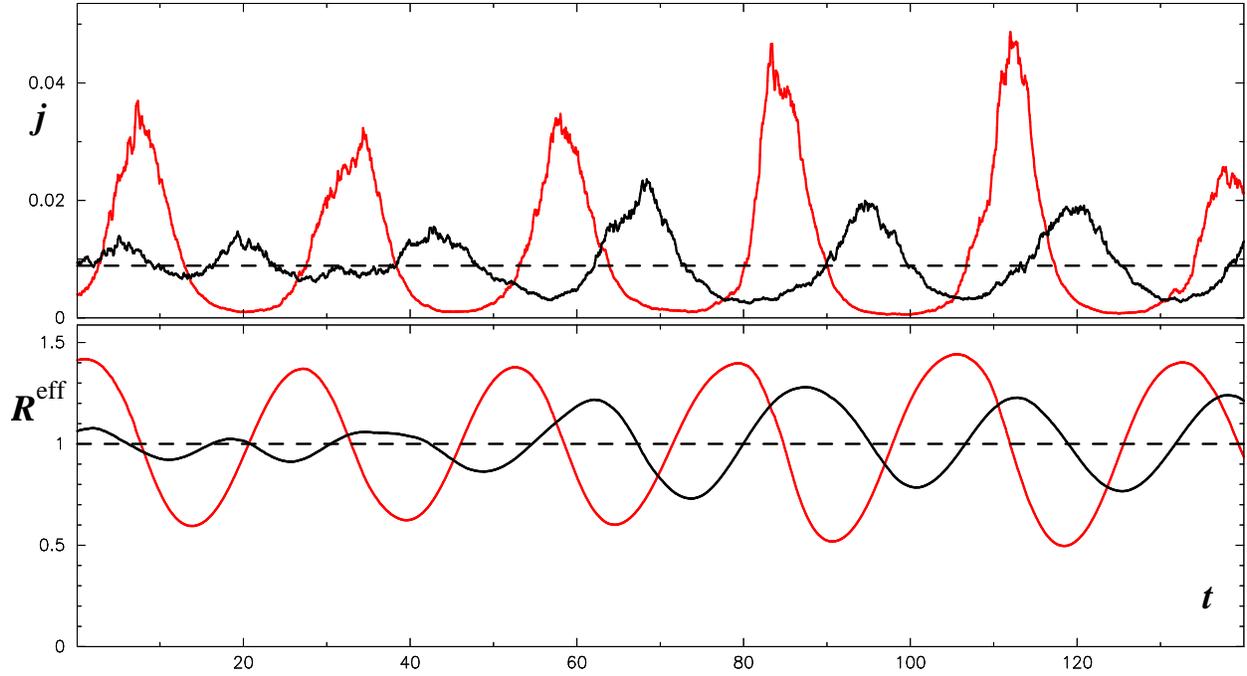}}  

  \caption{\label{fig7} Infection number and $R^{\mbox{\small eff}}$ for the stochastic
    Erlang kernel model with $\sigma=0.1$, other parameters as in fig.\ref{fig5}.}

\end{figure}

The same simulations for the Erlang kernel show a significant difference in the
behavior of the effective reproduction number, fig.\ref{fig7}. Due to the
integration over a continuous kernel, $R^{\mbox{\small eff}}$ is a smooth function
of $t$ and the fluctuations are only pronounced in $j(t)$. For both kernels, the
oscillations become much more irregular and the frequencies are distributed over
an area with width $\sim\sigma$.

\begin{figure}[!ht]

  \centerline{\includegraphics[width=0.7\tw]{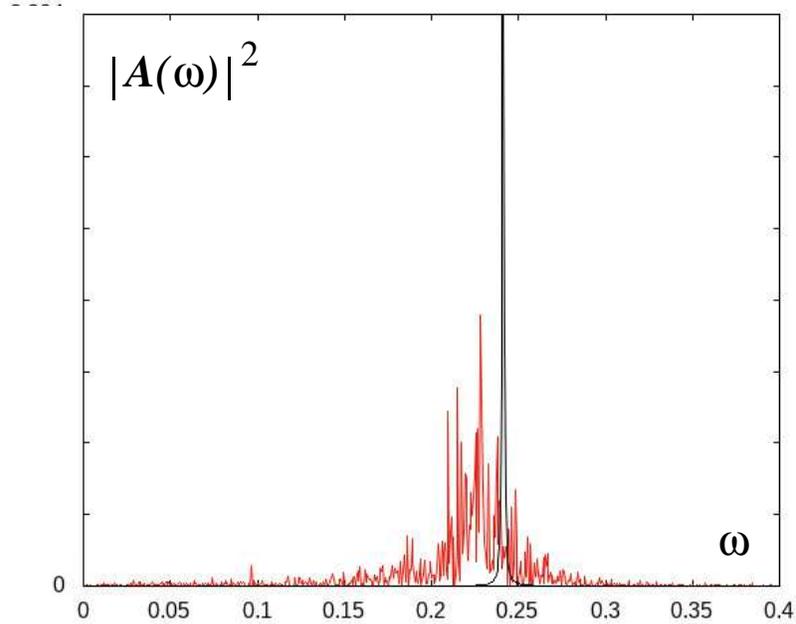}} 

  \caption{\label{fig8} Fourier transform $|A(\omega)|^2$ (arbitrary units)
    of a long time series $R^{\mbox{\small eff}}(t)$
    for $200<t<10000$ for the Erlang kernel with $\eta=7$ and
    $r_0=1.6,\ \alpha=50,\ \mu=0.1$. Black: $\sigma=0$, red: $\sigma=0.25$.} 

\end{figure}

In fig.\ref{fig8} we show the Fourier transform
\[ A(\omega_k) = \sum_n^N R^{\mbox{\small eff}}(n\delta t)
\exp\left(\frac{2\pi ink}{t_1-t_0}\right),\qquad \omega_k = 2\pi k/(t_1-t_0) \]
of a rather long time series up to $t_1-t_0=10000$, corresponding to about 400
oscillations. The function $R^{\mbox{\small eff}}$ is sampled with $N=2^{16}=65536$ points
with about 160 points per period.

\begin{figure}[!ht]

  \centerline{\includegraphics[width=\tw]{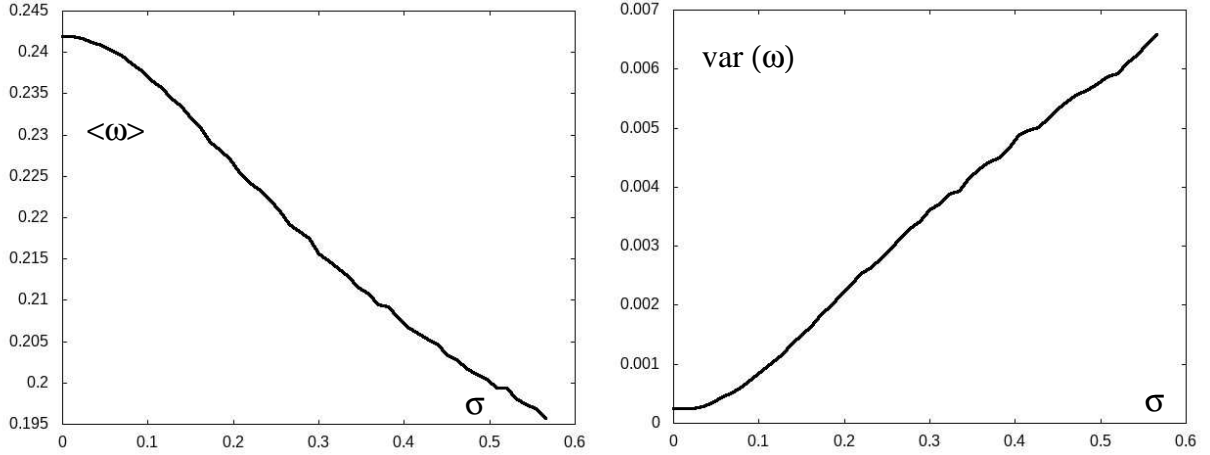}}   

  \caption{\label{fig9} Mean value (left) and variance of $\omega$ as function of
    $\sigma$ for the parameters of fig.\ref{fig8}. The data is an average over 30
    runs with the same parameters but different noise realizations.}

\end{figure}

\begin{figure}[!ht]

  \centerline{\includegraphics[width=\tw]{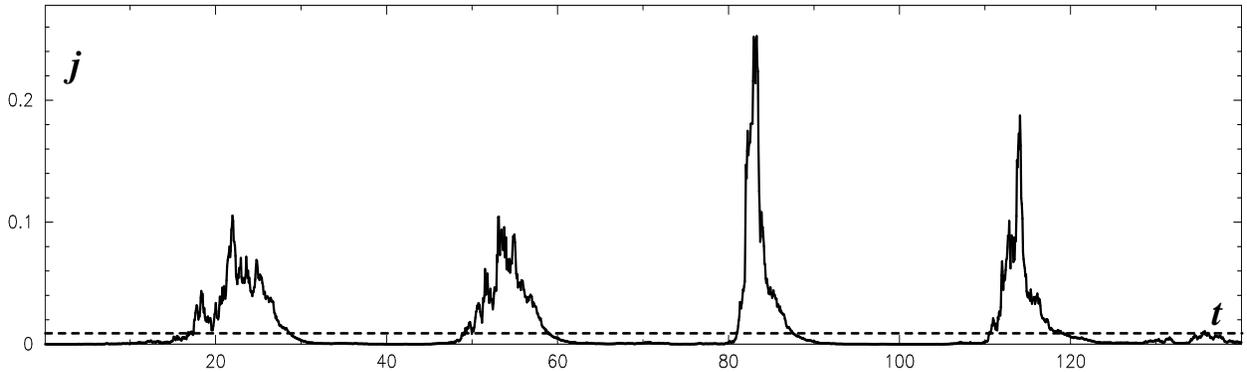}}  

  \caption{\label{fig10} Infection number for the stochastic
    Erlang kernel ($\eta=7$) with large fluctuations $\sigma=0.6$.}
  
\end{figure}

Fig.\ref{fig9} shows the mean frequency
\[ <\omega> = \frac{1}{N}\sum_k \omega_k\,|A(\omega_k)|^2 \]
and the variance
\[ \mbox{var}(\omega) = <\omega^2> - <\omega>^2 \]
over the fluctuation strength $\sigma$. It is clear that for rather large fluctuations the
oscillations become very irregular but the main frequency clearly persists,
fig.\ref{fig10}. We observe that the mean frequency decreases slightly with
increasing $\sigma$.

\subsubsection{Outbreaks}

For large $\sigma$ the infection dynamics shows long phases where the infection number
remains very small, interrupted by sharp periodic bursts, fig.\ref{fig10}. The amplitudes of
these outbreaks
are larger up to a factor 10 than those for the deterministic model (fig.\ref{fig5})
and may differ strongly from each other. In this context it is interesting to note
that for large $\sigma$, the minimal values for $j(t)$
become very small. For the series with $\sigma=0.6$ we have min$(j)\approx10^{-6}$,
for $\sigma=1$ we find min$(j)\approx10^{-9}$. But if the population $N$ is finite,
the minimal number of infected individuals according to (\ref{sir_def}) is
$I_m = N\,\mbox{min}(j)$. If we take $N\approx10^8$, corresponding to the population of a
rather large country, for $j<10^{-8}$ there would be no infected
individual anymore and the disease would have become extincted. Thus, large fluctuations could
lead to extinction even if the basic reproduction number stays larger than one,
a result already shown by Cai et al. [2015]. For our model we estimate the critical
sigma for extinction with
\[ \sigma_c = \frac{\sqrt{2(r_0-1)}}{r_0} \ . \]
Not very realistic if for instance compared with data from the
recent COVID waves are the rather equal times between the outbreaks.  Here it could
be possible to include fluctuations also in the immunity loss rate or in the delay
times of the feedback control. External effects like seasonal variations could be
included as well, a program that we want to study in forthcoming work.

\section{Conclusions}

In this paper we studied the influence of delayed feedback control on the dynamics
of a standard SIRS model. Delay terms normally generate oscillatory (Hopf) instabilities
of otherwise stable fixed points if the delay time exceeds a certain critical value.
Finite time delays, or, for the continuous case, memory effects come into play quite
naturally by the rather long time scales of macroscopic effects like a finite time of
immunity, the time necessary for the emergence of certain virus mutants, or the time
needed to establish mitigation measures. From this list it is clear that there exist
many possibilities to extend the model including one or even more memory terms with
different kernels. For an upcoming project it could be of interest to study the interplay
of two or more different delay terms on an otherwise low-dimensional deterministic
dynamics and see if quasi-periodic or even chaotic solutions may occur. It is known for
long that rather simple delay-differential equations like the Mackey-Glass equation
\cite{mack} or the sinusoidal nonlinearity \cite{sprott} show chaotic solutions if the
time delay becomes large enough.

Further interesting generalizations could be opened by considering fat-tailed memory
kernels with power-law decays and with diverging means (very long delay times).
Accounting for such kernels leads naturally to time-fractional generalizations of SIR
or SIRS models. In particular the combination of random walk models and memory effects
induced by renewal processes such as the fractional Poisson process and its generalizations 
\cite{MetzlerKlafter2000,SandevChechkinMetzler2021,mipo,granger} (and many others, see
references therein) may be of interest as well.

On the other hand, additional degrees of freedom may be encountered by including
environmental noise leading to fluctuating parameters of the SIR or SIRS models.
Then, simulations in the frame of stochastic nonlinear differential equations
with multiplicative noise come into the focus of attention.
The present paper tries to study the combined influence of retarded feedback control
and fluctuations due to a coupling to the environment onto the same parameter, namely the
infection rate. We found that noise my lead to large fluctuations of amplitude and
frequency of the otherwise very regular oscillations provided by the time delayed
feedback control alone.
In this context, the discussion of a corresponding Fokker-Planck equation of
(\ref{sirdst}) could be of interest. Such an equation was derived for
(\ref{sirdst}) in \cite{cai}, but for the case without delay terms. Here we would
need the extension of the Fokker-Planck theory to delay terms, a task that
we shall leave for future work.

Our model can be extended in different directions. A finite duration of being immune
can as well be included and modeled by a memory term with another given PDF as done in
our recent paper \cite{besmi}. Spatial effects can be taken into account by including
diffusion terms in the infection rate equation or considering the dynamics on small-world
networks. Finally, space and time varying infection rates could be introduced to model
seasonal variations and density distributions of the population. 


\end{document}